\documentclass[12pt]{article}  
\usepackage{cite}
\usepackage{epsfig}
\usepackage{graphicx}
\usepackage{amsmath}
\usepackage{amssymb}
\usepackage{ulem}
\usepackage{mdwlist}          
\usepackage{color}              

\usepackage{a41}
\usepackage{color}
\usepackage[rflt]{floatflt}
\usepackage{float}
\usepackage{slashed}

\usepackage{array}

\usepackage[english]{babel}
\usepackage[latin1]{inputenc}
\usepackage[T1]{fontenc}
\usepackage{ae}

\usepackage{url}

\usepackage{amsmath, amsthm, amssymb}
\newtheorem{thm}{Theorem}[section]

\newtheorem{definition}[thm]{Definition}

\newcommand{\Li}{{\rm Li}}

\newcommand{\Mvec}{{\rm\bf M}}

\usepackage{rotating}

\usepackage{graphicx}

\newcounter{mmacnt}
\def\restartmma{\setcounter{mmacnt}{0}}
\restartmma \catcode`|=\active
\def|#1|{\mathrm{#1}}
\catcode`|=12
\newenvironment{mma}{
 \par\smallskip
 \catcode`|=\active
 \parskip=0pt\parindent=0pt 
 \small
 \def\In##1\\{%
   \def\linebreak{\hfill\break\null\qquad}%
   \refstepcounter{mmacnt}
   \hangindent=2.5em\hangafter=0
   \leavevmode
   \llap{\tiny\sffamily In[\arabic{mmacnt}]:=\kern.5em}%
   \mathversion{bold}\footnotesize$\displaystyle##1$\normalsize
   \mathversion{normal}\par
 }%
 \def\Print##1\\{%
   \def\linebreak{		\hfill\break}%
   \hangindent=2.5em\hangafter=0
   \leavevmode ##1\par}%
 \def\Out##1\\{%
   \def\linebreak{$\hfill\break\null\hfill$}%
   \kern\abovedisplayskip\par
   \hangindent=2.5em\hangafter=0
   \leavevmode
   \llap{\tiny\sffamily Out[\arabic{mmacnt}]=\kern.5em}
   \footnotesize$\displaystyle##1$\normalsize\hfill\null\par
   \kern\belowdisplayskip
 }%
 \def\Warning##1##2\\{%
   \def\linebreak{\hfill\break}%
   \hangindent=2.5em\hangafter=0
   \leavevmode
   {\scriptsize##1 : ##2}\par}%
}{%
 \par\smallskip
}

\usepackage{color}

\newenvironment{fshaded}{%
\MakeFramed {\FrameRestore}
}%
{\endMakeFramed}

\allowdisplaybreaks[4]

\begin{document}
\setlength{\baselineskip}{0.515cm}
\sloppy
\thispagestyle{empty}
\begin{flushleft}
DESY 13--064
\hfill 
\\
DO--TH 17/12\\
September 2018\\
\end{flushleft}

\mbox{}
\vspace*{\fill}
\begin{center}

{\LARGE\bf Numerical Implementation of} 

\vspace*{3mm} 
{\LARGE\bf Harmonic Polylogarithms to Weight  w = 8}

\vspace{3cm}
\large
J.~Ablinger$^a$,
J.~Bl\"umlein$^b$, 
M.~Round$^{a,b}$, and
C.~Schneider$^a$, 

\vspace{1.cm}
\normalsize
{\it $^a$~Research Institute for Symbolic Computation (RISC),\\
                          Johannes Kepler University, Altenbergerstra\ss{}e 69,
                          A--4040 Linz, Austria}\\

\vspace*{3mm}

{\it  $^b$ Deutsches Elektronen--Synchrotron, DESY,}\\
{\it  Platanenallee 6, D-15738 Zeuthen, Germany}


\end{center}
\normalsize
\vspace{\fill}
\begin{abstract}
\noindent
We present the {\tt FORTRAN}-code {\tt HPOLY.f} for the numerical calculation of harmonic 
polylogarithms up to {\sf w = 8} at  an absolute accuracy of $\sim 4.9 \cdot 10^{-15}$ or better. 
Using algebraic and argument relations the numerical representation can be limited to the 
range $x \in [0, \sqrt{2}-1]$. We provide replacement files to map all harmonic polylogarithms 
to a basis and the usual range of arguments $x \in ]-\infty,+\infty[$ to the above interval 
analytically. We also briefly comment on a numerical implementation of real valued cyclotomic 
harmonic polylogarithms. 
\end{abstract}

\vspace*{\fill}
\noindent

\newpage
\section{Introduction}
\label{sec:1}

\vspace*{1mm}
\noindent
Many analytic higher order and multi-leg calculations result into representations based on harmonic 
polylogarithms ({\tt HPLs}) \cite{Gehrmann:2001ck,Vermaseren:2005qc,Ablinger:2014nga}. This function space has been 
introduced in Ref.~\cite{Remiddi:1999ew}. It is formed by the Volterra-iterative integrals over the 
alphabet
\begin{eqnarray}
f_{-1}(x) = \frac{1}{1+x},~~~~
f_{0}(x) = \frac{1}{x},~~~~
f_{1}(x) = \frac{1}{1-x},
\end{eqnarray}
with 
\begin{eqnarray}
H_{b,\vec{a}}(x) &=& \int_0^x dy f_b(y) H_{\vec{a}}(y),~~~~~H_\emptyset = 1,
\\
H_{\underbrace{\mbox{\scriptsize 0, \ldots ,0}}_{n}}(x) &=& \frac{1}{n!} H^n_0(x).  
\end{eqnarray}
By definition all {\tt HPLs} are zero for $x=0$.
The harmonic polylogarithms are generalizations of the classical polylogarithms \cite{DILOG4,DUDE}
and the Nielsen integrals \cite{NIELSEN1,NIELSEN2,NIELSEN3}. They obey algebraic \cite{Blumlein:2003gb} 
and argument relations \cite{Remiddi:1999ew}. Furthermore, they are dual structures, 
by the Mellin transform
\begin{eqnarray}
\Mvec[f(x)](N) = \int_0^1 dx x^{N-1} f(x),
\end{eqnarray}
to the nested harmonic sums \cite{Vermaseren:1998uu,Blumlein:1998if} 
\begin{eqnarray}
S_{b,\vec{a}}(N) = \sum_{k=1}^N \frac{({\rm sign}(b))^k}{k^{|b|}} S_{\vec{a}}(k),~~S_\emptyset = 1,~~~~b, a_i \in 
\mathbb{Z} \setminus \{0\}, N \in \mathbb{N} \backslash \{0\}.
\end{eqnarray}
In the past numerical representations of a series of classical polylogarithms and Nielsen integrals were given
in \cite{NIELSEN2} and have been in wide use for decades. The generalization to harmonic polylogarithms 
has been performed in \cite{Gehrmann:2001pz} up to weight {\sf w = 4} and in a private version to {\sf w = 5}.
An implementation using the package {\tt Ginac} \cite{Bauer:2000cp} has been given in \cite{Vollinga:2004sn}.
The {\tt Mathematica} packages {\tt HPL} \cite{Maitre:2005uu} 
and {\tt HarmonicSums} \cite{HARMONICSUMS,Ablinger:PhDThesis,Ablinger:2011te,Ablinger:2013cf,Ablinger:2014bra} 
 do also allow the numerical
evaluation of harmonic polylogarithms. There is also an implementation in {\tt Maple} \cite{Frellesvig:2018lmm}.
In various applications one needs the numerical representation of higher weight harmonic polylogarithms,
e.g. to weight {\sf w = 6} in Refs.~\cite{Moch:2017uml,Ablinger:2018yae,FORMF2}, which were given in 
\cite{Ablinger:2017tqs}. Numerical implementations of the analytic continuation of harmonic sums to complex 
arguments have been given in \cite{ANCONT}.

In the foreseeable time fast numerical implementations of harmonic polylogarithms are also needed for higher
weights. In the present paper we present a {\tt FORTRAN} implementation for harmonic polylogarithms up to 
weight {\sf w = 8}. If one compares the computational times between {\tt FORTRAN} implementations and 
the ones in {\tt Ginac} \cite{Vollinga:2004sn} or {\tt Mathematica} \cite{Maitre:2005uu,HARMONICSUMS} the 
latter ones are somewhat slower. In many phenomenological and experimental applications users prefer to work 
with numerical programmes only. The other existing programmes have of course also advantages against pure
numeric implementations, since they can be tuned to arbitrary precision and allow analytic functional mappings.

The number of 
harmonic polylogarithms grows rapidly with the weight. Therefore it is necessary to give numerical 
representations only after the algebraic relations have been used. Furthermore, it is sufficient to represent
the harmonic polylogarithms in the core interval $x \in [0, \sqrt{2}-1]$ since the representations for the 
remaining real domain can be given by functional relations.

In Section~\ref{sec:2} we describe the numerical implementation for the harmonic polylogarithms. Technical 
aspects on the code and its use are discussed in Section~\ref{sec:3}. In Section~\ref{sec:4} we comment
on the corresponding numerical implementation for cyclotomic harmonic polylogarithms\footnote{
Beyond these functions various more special functions contribute in analytic calculations in Quantum Field 
Theories. For recent surveys see \cite{Ablinger:2013jta,Blumlein:2018cms}.}
and Section~\ref{sec:5}
contains the conclusions. Numerical examples are given in Appendix~A. Appendix~B contains the list of
auxiliary files attached to this paper to perform preparatory calculations in {\tt Mathematica}. Others are 
subsidiary files used in the initialization of the {\tt 
FORTRAN}-code.
\section{The Numerical Implementation}
\label{sec:2}

\vspace*{1mm}
\noindent
In order to keep the numerical implementation as compact as possible the user is asked to perform mappings on his
expressions beforehand into the main interval $x \in [0,\sqrt{2}-1]$. This can be done using the {\tt Mathematica} 
package {\tt HarmonicSums} 
\cite{HARMONICSUMS,Ablinger:PhDThesis,Ablinger:2011te,Ablinger:2013cf,Ablinger:2014bra}. For 
this purpose we work with bases of harmonic polylogarithms for the different weights. 
These are listed in the files {\tt LIST1.m, ..., LIST8.m} attached to this paper. The file {\tt LIST.m} 
contains the combined basis.
The number of basis elements 
per weight is given in Table~\ref{TAB1}.

\begin{table}[H]\centering
\renewcommand{\arraystretch}{1.3}
\begin{tabular}{|r|r|}
\hline \hline
\multicolumn{1}{|c|}{weight} & \multicolumn{1}{c|}{\# of basis elements}\\
\hline
   1  &   3    \\
   2  &   3    \\
   3  &   8    \\
   4  &  18    \\
   5  &  48    \\
   6  & 116    \\
   7  & 312    \\
   8  & 810    \\
\hline \hline
\end{tabular}
\caption[]{\sf Number of basis elements as a function of the weight.}
\label{TAB1}
\renewcommand{\arraystretch}{1.0}
\end{table}

\noindent
The number of basis elements can be calculated using a formula by Witt, cf.~\cite{Blumlein:2003gb}, for a 
3-letter alphabet
\begin{eqnarray}
N_{\rm bas}({\sf w}) = \frac{1}{\sf w} \sum_{d|w} \mu\left(\frac{w}{d}\right) 3^d. 
\end{eqnarray}
Here $\mu$ denotes the M\"obius function.

We have e.g.
\begin{eqnarray}
{\tt LIST1.m} &=& \{H[1,x], H[0,x], H[-1,x]\}
\\
{\tt LIST2.m} &=& \{H[0,1,x], H[-1,1,x], H[-1,0,x]\}
\\
{\tt LIST3.m} &=& \{
H[0,1,1,x], 
H[0,0,1,x], 
H[-1,1,1,x], 
H[-1,1,0,x], 
H[-1,0,1,x], 
H[-1,0,0,x],
\nonumber\\ && 
H[-1,-1,1,x],
H[-1,-1,0,x]\}.
\end{eqnarray}
The relations of all other {\tt HPLs} are stored in the replacement list
\begin{eqnarray}
{\tt hrel.m}~~~~~[12.8~{\rm Mbyte}]
\end{eqnarray}
attached to this paper.

Usually one has expressions with main argument $x$ of the harmonic polylogarithms in $x \in ]-\infty,+\infty[$.
For the arguments $x \in ]-\infty,0]$ one first applies the procedure 
\begin{eqnarray}
{\tt TransformH}[H_{\vec{a}}(f(x)),x]
\end{eqnarray}
of {\tt HarmonicSums}.

\noindent
\underline{Example:}
\begin{eqnarray}
\hspace*{-5mm}
{\tt TransformH}[H[-1,0,0,-x],x] &=& -H[1,x] H[0,0,-x] + H[0,-x] H[0,1,x] - H[0,0,1,x]
\nonumber\\ 
\hspace*{-5mm}
&=& - \frac{1}{2} H[1,x] (H[0,x] + i \pi)^2 + (H[0,x] + i \pi)  - H[0,0,1,x]. 
\nonumber\\
\end{eqnarray}

\noindent
The same command maps $x \in [1,\infty]$ to $x \in [0,1]$. 
\newline
\underline{Example:}
\begin{eqnarray}
{\tt TransformH}\left[H\left[-1,0,0,\frac{1}{x}\right]\right] &=& H[-1,0,0,x] - \frac{1}{6}H[0,x]^3.
\end{eqnarray}
Now only two regions have to be considered:
\begin{eqnarray}
x \in [0, \sqrt{2} - 1]~~~~\text{and}~~~~x \in [\sqrt{2} - 1,1].
\label{eq:inte}
\end{eqnarray}
The expression in the second region is obtained from the first one by the mapping
\begin{eqnarray} x \mapsto \frac{1-x}{1+x}.
\end{eqnarray}
\underline{Example:}
\begin{eqnarray}
{\tt TransformH}\left[H\left[-1,0,0,\frac{1-x}{1+x}\right]\right] &=& \frac{3}{4} \zeta_3
- \frac{1}{6} H[-1,x]^3 - H[-1,x] H[-1,1,x] 
\nonumber\\ && + H[-1,-1,1,x] 
- H[-1,1,1,x].
\end{eqnarray}
In the case one is leaving the basis one uses 
\begin{eqnarray}
{\tt ReduceToHBasis}[{\it expression}]
\end{eqnarray}
to get back to the basis again. In order to carry out these function calls, the following 
subsidiary files are useful. More precisely,
the corresponding substitution rules are tabulated in the file {\tt hrel.m}.
For the other operations  we provide the following substitution files :
\begin{align}
{\tt LISTOMX.m}~~&~~x \mapsto -x               & [~~2.6~{\rm Mbyte}] \\
{\tt LISTOOX.m}~~&~~x \mapsto \frac{1}{x}      & [13.1~{\rm Mbyte}] \\
{\tt LISTMTX.m}~~&~~x \mapsto \frac{1-x}{1+x}  & [13.7~{\rm Mbyte}]
\end{align}

\noindent
In case we have {\tt HPLs} over a two-letter alphabet, 
i.e. $\{0,-1\}$ or $\{0,1\}$, only one branch cut is present, i.e. $[-1,-\infty[$ or $[1,\infty[$.
In all other cases one has to consider a cut separated representation of the respective {\tt HPL}. 
This cut separation is performed by using the command
\begin{eqnarray}
{\tt HSeparateCuts}[H_{\vec{a}}(x)].
\end{eqnarray}
\underline{Example:}
\begin{eqnarray}
{\tt HSeparateCuts}[H[-1,1,1,x]] &=& \{{\tt term1, term2}\}
\nonumber\\ &=& \left\{ -\frac{1}{2} \ln^2(2) H[-1,x] + H[-1,1,1,x], \frac{1}{2} \ln^2(2) H[-1,x] \right\}.
\nonumber\\
\end{eqnarray}
The sum of the two terms yields the original expression again. 

To reduce this representation in optimal terms one also needs the replacement table for the multiple zeta values up 
to
{\sf w=8} \cite{Blumlein:2009cf}. All the polynomial basis constants contribute,
\begin{eqnarray}
&&\Bigl\{
\ln(2),
\zeta_2, 
\zeta_3, 
\Li_4\left(\frac{1}{2}\right), 
\zeta_5, 
\Li_5\left(\frac{1}{2}\right),
\Li_6\left(\frac{1}{2}\right),
s_6,
\zeta_7,
\Li_7\left(\frac{1}{2}\right),
s_{7a}, s_{7b},
\Li_8\left(\frac{1}{2}\right),
\nonumber\\ &&
s_{8a}, s_{8b}, s_{8c}, s_{8d}
 \Bigr\},
\end{eqnarray}
with
\begin{eqnarray}
\zeta_k &=& S_{k}(\infty),~~~k \geq 2
\\
\Li_k\left(x\right) &=& \sum_{l=1}^\infty \frac{x^l}{l^k}
\\
s_6    &=&  S_{-5,-1}(\infty)
\\ 
s_{7a} &=&  S_{-5,1,1}(\infty)  
\\ 
s_{7b} &=&  S_{5,-1,-1}(\infty)
\\ 
s_{8a} &=&  S_{5,3}(\infty)
\\ 
s_{8b} &=&  S_{-7,-1}(\infty)
\\ 
s_{8c} &=&  S_{-5,-1,-1,-1}(\infty)
\\ 
s_{8d} &=&  S_{5,-1,1,1}(\infty).
\end{eqnarray}

Now the Bernoulli-improvement \cite{tHooft:1978jhc}, see also \cite{Gehrmann:2001pz},
of the respective series expansion is performed. 
To this end we substitute
\begin{eqnarray}
{\tt term1(x)} &\mapsto& {\tt term1(x)}|_{x \rightarrow 1 - e^{-v}} \\
{\tt term2(x)} &\mapsto& {\tt term1(x)}|_{x \rightarrow e^{u} - 1}, 
\end{eqnarray}
with 
\begin{eqnarray}
u &=& \hspace*{4mm} \ln(1+x)\\
v &=& -\ln(1-x)
\end{eqnarray}
and expand into series in $u$ and $v$ up to a certain order $N_{max}$, which we 
choose $N_{max} = 40$. This representation converges much faster than the corresponding Taylor series.
The representations obtained are power series in $u$ and $v$ and logarithms of $u$ and $v$ to a certain power. 
At weight {\sf w} there are contributions of up to $O(\ln^{\sf w-1}(u))$ and of $O(\ln^{\sf w-2}(v))$.
In Appendix~\ref{APPa} we give one example for illustration.

One stores now the corresponding constants appearing in the expansion for each weight {\sf w} in a data file.
These constants are read into a very compact {\tt FORTRAN} programme. Herewith we obtain a representation of 
all basis {\tt HPLs} up to {\sf w = 8} for $x \in \mathbb{R}$, after arguments from all other regions are 
mapped using the commands described above. This allows to work with real representations only. Complex 
valued {\tt HPLs} are already separated into their real and imaginary part by the above transformations.
The corresponding auxiliary files can be used in an usual {\tt Mathematica} session even without invoking
the package {\tt HarmonicSums}.
\section{Description of the code \label{sec:3}}

\vspace*{1mm}
\noindent
The code {\tt HPOLY} provides the calculation of all {\tt HPLs}, using the basis representation and
argument mappings described in Section~\ref{sec:2} in the region $x \in [0, \sqrt{2}-1]$. It shall be 
compiled using 
\[{\tt gfortran~~hpoly.f}.\] 
\noindent
in {\tt LINUX}. There is a user-routine {\tt UHPOLY} through which the 
corresponding calculations can be made. Through the initialization routine {\tt UHPOLYIN}
the maximal weight of the calculation is chosen setting the variable {\tt IW}. If the calculation of an {\tt HPL} is 
intended at a higher weight than {\sf w = 8}, the code cannot perform the calculation. Choosing {\tt IW = 8} the 
initialization time
is longest with 0.2375 sec for reading in the constant tables. 
Through {\tt UHPOLY} the user can arrange special output formats. Only the routines {\tt UHPOLYIN}
and {\tt UHPOLY} are free to be modified by the user.

A built in logic checks whether the required {\tt HPL} is contained in the basis and whether the 
variable is in the correct range. The routine {\tt HPOLYIN} performs necessary initializations.

The main routine is
\begin{verbatim}      
                PROGRAM HPOLY
          *          
          *--------------------------------------------------------------    
          *   
          *     numerical evaluation of HPLs up to weight w = 8  
          *   
          *     J. Bluemlein, 17.9.2018   
          *   
          *--------------------------------------------------------------    
          *
                WRITE(6,*) 
               &'**** HPOLY: calculation of HPLS up to w=8 ****' 
                CALL UHPOLYIN
          * 
                CALL HPOLYIN
                WRITE(6,*) '**** HPOLY: initialization finished ****' 
          *
                CALL UHPOLY
          *  
                WRITE(6,*) '**** HPOLY: finished ****' 
                STOP
                END
                SUBROUTINE UHPOLYIN
          * 
          *--------------------------------------------------------------    
          * 
                IMPLICIT NONE
                INTGER   IW 
                COMMON/WEIGHT/ IW
          *
                IW = 8
          * 
                RETURN
                END  
                SUBROUTINE UHPOLY
          * 
          *--------------------------------------------------------------    
          * 
                IMPLICIT NONE
          *
                REAL*8   H1,H2,H3,H4,H5,H6,H7,H8
                REAL*8   T(8),X
                INTEGER  K
          *  
                EXTERNAL H1,H2,H3,H4,H5,H6,H7,H8  
          *
                CALL CPU_TIME(START)
                X   =  0.3D0
                CALL   HPOLYC(X)
          * 
                T(1)  =  H1(-1,X) 
                T(2)  =  H2(0,1,X) 
                T(3)  =  H3(-1,0,0,X) 
                T(4)  =  H4(-1,-1,1,0,X) 
                T(5)  =  H5(-1,-1,1,0,1,X)
                T(6)  =  H6(-1,0,-1,1,1,1,X)
                T(7)  =  H7(-1,-1,1,1,0,1,0,X)
                T(8)  =  H8(-1,0,-1,0,-1,0,1,1,X)
          *
                DO 1 K = 1,8
                WRITE(6,*) 'K, X , T=', K,X,T(K)
          1     CONTINUE
                CALL CPU_TIME(END)
                WRITE(6,*) 'CPU TIME=', END-START, '   SEC'
  * 
                RETURN
                END  
          *--------------------------------------------------------------    
\end{verbatim}
Running the above example produces the following output values:
\begin{table}[H]\centering
\renewcommand{\arraystretch}{1.5}
\begin{tabular}{|l|r|r|}
\hline \hline
\multicolumn{1}{|c|}{{\tt HPL}} 
& \multicolumn{1}{c|}{value}
& \multicolumn{1}{c|}{accuracy}
\\
\hline
  {\tt H[-1,x]}                 &    0.26236426446749106e-0 &    7.96e-18  \\
  {\tt H[0,1,x]}                &    0.32612951007547608e-0 &    1.05e-17  \\           
  {\tt H[-1,0,0,x]}             &    0.81699704232693138e-0 &    1.20e-16  \\ 
  {\tt H[-1,-1,1,0,x]}          & $-$0.10536957058865759e-0 &    1.20e-16  \\ 
  {\tt H[-1,-1,1,0,1,x]}        &    0.27247014022231675e-3 &    3.97e-17  \\
  {\tt H[-1,0,-1,1,1,1,x]}      &    0.45411840144185533e-5 &    8.46e-19  \\
  {\tt H[-1,-1,1,1,0,1,0,x]}    & $-$0.80698691040978487e-4 &    3.86e-18  \\
  {\tt H[-1,0,-1,0,-1,0,1,1,x]} &    0.57153046109648109e-6 &    3.47e-18  \\ 
\hline \hline
\end{tabular}
\renewcommand{\arraystretch}{1.0}
\caption[]{\sf The result of a test calculation of the code {\tt HPOLY.f} with $x = 0.3$.}
\label{TAB5}
\end{table}

\noindent
The absolute accuracy has been determined comparing with the corresponding result obtained
using {\tt Ginac} \cite{Vollinga:2004sn}. We have compared the numerical values produced by {\tt HPOLY} and
using {\tt Ginac} forming the difference in a {\tt Maple} programme, which can handle floating point number
inputs straightforwardly, seeming to be more subtle in {\tt Mathematica}.
\section{Cyclotomic Harmonic Polylogarithms \label{sec:4}}

\vspace*{1mm}
\noindent
In the following we would like to make a few brief remarks on the numerical evaluation of cyclotomic harmonic 
polylogarithms \cite{Ablinger:2011te}.

These functions emerge in massive calculations from 3-loops onward. Usually their main argument $x$ is located in 
the interval $[-1,1]$. This is going to be the range we are considering in the following. The new letters are of 
cyclotomy 3,4 and 6~:
\begin{eqnarray}
\left\{
\frac{1}{1+x+x^2},~~
\frac{x}{1+x+x^2},~~
\frac{1}{1+x^2},~~
\frac{x}{1+x^2},~~
\frac{1}{1-x+x^2},~~
\frac{x}{1-x+x^2}\right\}.
\end{eqnarray}
While here the transformations $x \mapsto 1/x$ and $x \mapsto -x$ are class preserving, the transformation 
$x \mapsto (1-x)/(1+x)$ is not. There is also no reason, why the above Bernoulli-improvement should accelerate the 
convergence of the expansions, given the structure of the above letters.

Yet one would like to have a series representation in the interval $[-1,1]$, see e.g. 
Ref.~\cite{Ablinger:2018yae,FORMF2}. One possibility is given by Taylor series in $x$ around $x=0, 1$ and $x=-1$,  
which can be calculated analytically using the command 
\begin{eqnarray}
{\tt HarmonicSumsSeries[f(x),\{x,0,n\}]}
\end{eqnarray}
of {\tt HarmonicSums}, where {\tt n} denotes the highest power in the expansion. The series are
intended to hold at a convergence radius of $\rho = 1/2$ at double precision, i.e. $53 \cdot 
\ln_{10}(2) \approx 15.955$ digits \cite{DP}. 
The expansion around $x = -1$ may be performed by mapping 
\begin{eqnarray}
\left\{
\frac{1}{1+x+x^2} \leftrightarrow \frac{1}{1-x+x^2},~~
\frac{x}{1+x+x^2} \leftrightarrow - \frac{x}{1-x+x^2}
\right\}
\end{eqnarray}
more effectively. We note that some of the cyclotomic {\tt HPLs} can be complex for $x \in [-1,0[$.
\begin{table}[H]\centering
\renewcommand{\arraystretch}{1.5}
\begin{tabular}{|r|r|r|r|r|}
\hline \hline
\multicolumn{1}{|c|}{$x$} 
& \multicolumn{1}{c|}{{\tt HPL}}
& \multicolumn{1}{c|}{run time {\tt Fortran}}
& \multicolumn{1}{c|}{run time {\tt Ginac}}
& \multicolumn{1}{c|}{accuracy}
\\
\multicolumn{1}{|c|}{} 
& \multicolumn{1}{c|}{}
& \multicolumn{1}{c|}{[sec]}
& \multicolumn{1}{c|}{[sec]}
& \multicolumn{1}{c|}{}
\\
\hline
   $-$1.0  &  $-$0.19456824628601374558e+0   & 6.0e-6  & 2.4e-4 & 5.5e-18   \\    
   $-$0.8  &  $-$0.10723997730030421271e+0   & 6.0e-6  & 2.9e-4 & 4.5e-18   \\  
   $-$0.4  &  $-$0.16724871547938951751e$-$1 & 8.0e-6  & 3.3e-4 & 5.2e-18   \\
      0.4  &  ~~0.25064429005154578702e$-$1  & 8.0e-6  & 2.7e-4 & 2.3e-18   \\
      0.8  &  ~~0.19221120777067257053e+0    & 8.0e-6  & 6.2e-3 & 6.0e-18   \\
      1.0  &  ~~0.34039020776540250510e+0    & 2.9e-5  & 6.0e-3 & 1.5e-17   \\
\hline \hline
\end{tabular}
\renewcommand{\arraystretch}{1.0}
\caption[]{\sf Example of the numerical calculation of an cyclotomic {\tt HPL}~~{\tt H[\{6,1\},0,-1,x]} comparing 
the
run times in {\tt FORTRAN} and {\tt Ginac}.}
\label{TAB3}
\end{table}

\noindent

The accuracy of the respective representation can be 
tested 
comparing the numerical 
result with the complex-valued representations provided by {\tt Ginac}, cf.~Ref.~\cite{Vollinga:2004sn}.
To have these explicit series representations is of importance since they usually perform faster than
the algorithm based on a H\"older convolution in \cite{Vollinga:2004sn}. However, the latter one 
has the advantage that it can be extended to arbitrary precision.

\begin{figure}[H] 
\centering
\includegraphics[width=0.48\textwidth]{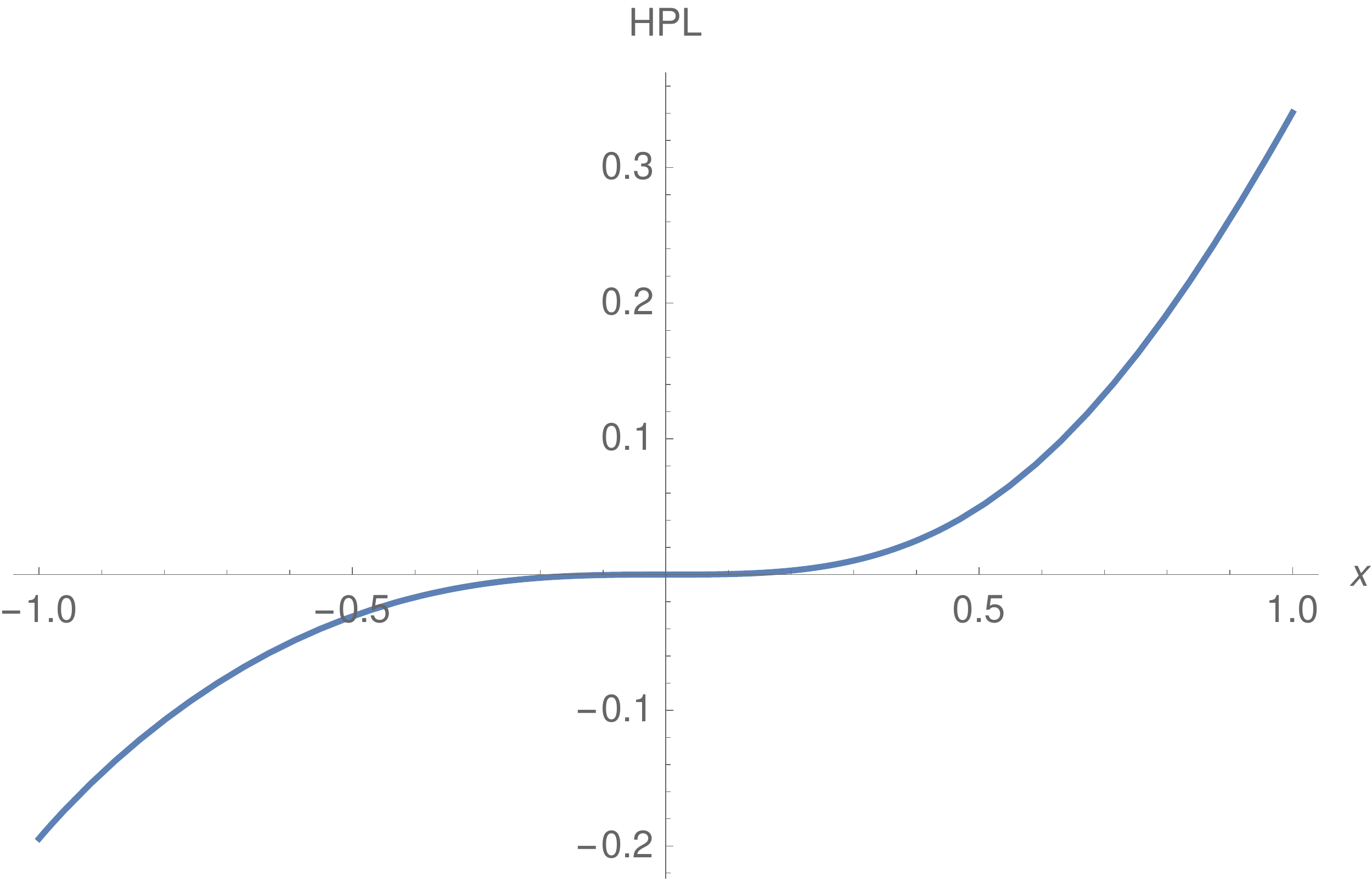}
\caption{\sf \small The function {\tt H[\{6,1\},0,-1,x]} in the region $x \in [-1,1]$.}
\label{fig2}
\end{figure}

In Table~\ref{TAB3} we summarize the performance for a series of values using both implementations by considering
the example of {\tt H(\{6,1\},0,-1,x)}. The function is graphically illustrated in Figure~\ref{fig2}.

\noindent
In case of the planar contributions to the 3-loop massive Formfactors \cite{Ablinger:2018yae,FORMF2} 206
real-valued cyclotomic polylogarithms up to {\sf w=6} contribute. Their numerical representation is given in
Ref.~\cite{FORMF2}.  
\section{Conclusions \label{sec:5}}

\vspace*{1mm}
\noindent
We presented a compact {\tt FORTRAN} code which calculates the {\tt HPLs} numerically up to weight {\sf w~=~8}. 
Computer-readable replacement files are provided to allow the analytic representation of a corresponding polynomial 
of {\tt HPLs} over the chosen algebraic basis, mapping the emerging arguments $x \in \mathbb{R}$ into the interval
$x \in [0, \sqrt{2} - 1]$ in a {\tt Mathematica} session during a preparatory step. For the implementation of the 
{\tt HPLs} we used the Bernoulli-improvement \cite{tHooft:1978jhc}. The numerical accuracy of the representation is 
$\sim 
4.9 \cdot 10^{-15}$ or better. We also compared run times with corresponding implementations in {\tt Ginac}. 
Because we limit the accuracy to about double precision, the {\tt FORTRAN} implementation is faster. This is of 
advantage for larger scale phenomenological and experimental applications. The present code extends earlier work of 
the authors of Ref.~\cite{Gehrmann:2001pz} to far higher weight.

We briefly commented on the numerical implementation of cyclotomic harmonic polylogarithms, which are important 
in massive higher loop calculations, as e.g. for the massive Formfactors from 3-loop order 
\cite{Ablinger:2018yae,FORMF2,STEINH}. Here, precise Taylor series around $x = 0, 1, -1$ providing sufficient 
convergence in a circle of radius $\rho = 1/2$ within $x \in [-1,1]$ serve the purpose. Detailed representations are
given in Ref.~\cite{FORMF2}.

The use of the {\tt FORTRAN}-code {\tt HPOLY.f} and of the attached subsidiary files requires the citation of the 
present paper. Discovered bugs shall be reported to {\tt Johannes.Bluemlein@desy.de}.

\appendix
\section{Numerical examples \label{APPa}}

\vspace*{1mm}
\noindent
\subsection{Harmonic Polylogarithms \label{APPa1}}

\vspace*{1mm}
\noindent
In the following we show the explicit representation for one example, {\tt H[-1,1,0,x]} and give some
numerical illustration. This harmonic polylogarithm is represented by
\begin{eqnarray}
\lefteqn{{\tt H[-1,1,0,x]}}   \nonumber\\ &&
= 1.51561421398405852765343970478 u -0.173286795139986327354308030365 u^2
\nonumber\\ &&
-9.62704417444368485301711279803e\text{-}3 u^3 + 4.81352208722184242650855639902e\text{-}5 u^5
\nonumber\\ &&
-5.45750803539891431576933832088e\text{-}7 u^7
+7.95886588495675004383028505128e\text{-}9 u^9
\nonumber\\ &&
-1.31551502230690083369095620682e\text{-}10 u^{11}
+2.34789839498327069523925944587e\text{-}12 u^{13}
\nonumber\\ &&
-4.41717498332597960030540850887e\text{-}14 u^{15}
+8.63804169263087598091731324426e\text{-}16 u^{17}
\nonumber\\ &&
-1.74017681474319135406058068735e\text{-}17 u^{19}
+3.58929549165142423790521910252e\text{-}19 u^{21}
\nonumber\\ &&
-7.5465410635862794583439948632e\text{-}21 u^{23}
+1.6120829755323789347574740235e\text{-}22 u^{25}
\nonumber\\ &&
-3.49013136556204594004342294876e\text{-}24 u^{27}
+7.64298689734656581222853583268e\text{-}26 u^{29}
\nonumber\\ &&
-1.69034929890524597704956962836e\text{-}27 u^{31}
+3.77081955969303929350628544829e\text{-}29 u^{33}
\nonumber\\ &&
-8.47603948488133928506876824739e\text{-}31 u^{35}
+1.91812087138166865720209408063e\text{-}32 u^{37}
\nonumber\\ &&
-4.36689003100059767808970725046e\text{-}34 u^{39}
\nonumber\\ &&
-1.51561421398405852765343970478 v
+2.45753828564099545590515613688e\text{-}1 v^2
\nonumber\\ &&
-7.28443369411397782763491496943e\text{-}3 v^3
-4.665222471648288365706911026793e\text{-}3 v^4
\nonumber\\ &&
+7.18180060085976020176571697142e\text{-}4 v^5
+1.0998660373605051065035495605e\text{-}5 v^6
\nonumber\\ &&
-1.15456136926644459648518129105e\text{-}5 v^7
+2.2668351288044216255466542194e\text{-}7 v^8
\nonumber\\ &&
+2.10153210137148470728310501335e\text{-}7 v^9 
-1.14687914794188966052447108046e\text{-}8 v^{10}
\nonumber\\ && 
-3.89908427576215576063242315571e\text{-}9 v^{11}
+3.66138107063939833413172284743e\text{-}10 v^{12}
\nonumber\\ && 
+7.06461980894443084509781017833e\text{-}11 v^{13}
-1.01701961998442644609478487375e\text{-}11 v^{14}
\nonumber\\ && 
-1.20382742279859366919809526002e\text{-}12 v^{15}
+2.62615442112180864875738033e\text{-}13    v^{16}
\nonumber\\ && 
+1.80691839540417471780902160508e\text{-}14 v^{17}
-6.45036155764417783408860256621e\text{-}15 v^{18}
\nonumber\\ && 
-1.94331333196745041013938386874e\text{-}16 v^{19}
+1.52063344250754104057653980337e\text{-}16 v^{20}
\nonumber\\ && 
-5.82510897939787787036716572212e\text{-}19 v^{21}
-3.44907968543220341667785192179e\text{-}18 v^{22}
\nonumber\\ && 
+1.26829440977979122960207894152e\text{-}19 v^{23}
+7.51151317182163069584552487431e\text{-}20 v^{24}
\nonumber\\ && 
-5.40148573489695792030638540445e\text{-}21 v^{25}
-1.55998517649156934189241423249e\text{-}21 v^{26}
\nonumber\\ && 
+1.76849481787857125449286908312e\text{-}22 v^{27}
+3.04523398314277895864778604675e\text{-}23 v^{28}
\nonumber\\ && 
-5.13172809239911634765975392106e\text{-}24 v^{29}
-5.41954452181938287426807053132e\text{-}25 v^{30}
\nonumber\\ && 
+1.38155545866828806840477032923e\text{-}25 v^{31}
+8.13686152516356159595967623328e\text{-}27 v^{32}
\nonumber\\ && 
-3.51809188882467289354219400405e\text{-}27 v^{33}
-7.47063186973607435240189362655e\text{-}29 v^{34}
\nonumber\\ && 
+8.54615019188533962991092101997e\text{-}29 v^{35}
-1.08160378090075874429236755202e\text{-}30 v^{36}
\nonumber\\ && 
-1.98536129319620756482783417757e\text{-}30 v^{37}
+9.18943710177493828604598228192e\text{-}32 v^{38}
\nonumber\\ && 
+4.40086958481705100275797309929e\text{-}32 v^{39}
-3.65462415959228934623749630635e\text{-}33 v^{40}
\nonumber\\ && 
+ \Bigl[
 0.693147180559945309417232121458 u      + 0.500000000000000000000000000000 u^2
\nonumber\\ &&
+0.333333333333333333333333333333 u^3    + 0.250000000000000000000000000000 u^4
\nonumber\\ &&
+0.216666666666666666666666666667 u^5    + 0.208333333333333333333333333333 u^6
\nonumber\\ &&
+0.214682539682539682539682539683 u^7    + 0.232291666666666666666666666667 u^8
\nonumber\\ &&
+0.260653659611992945326278659612 u^9    + 0.300834986772486772486772486772 u^{10}
\nonumber\\ &&
+0.355101661455828122494789161456 u^{11} + 0.426919505070546737213403880071 u^{12}
\nonumber\\ &&
+0.521158552417232972788528344084 u^{13} + 0.644462450647346480679814013147 u^{14}
\nonumber\\ &&
+0.805794410709134584795960457336 u^{15} + 1.01720115026784328619646079964  u^{16}
\nonumber\\ &&
+1.29486269565703958818843661670  u^{17} + 1.66052621302802363001326846124  u^{18}
\nonumber\\ &&
+2.14346104513832728152640246783  u^{19} + 2.78312455814376537110798639484  u^{20}
\nonumber\\ &&
+3.63279999838074563025362227705  u^{21} + 4.76456594745939637957111115559  u^{22}
\nonumber\\ &&
+6.27609256313695770804541041481  u^{23} + 8.29994698224797632624920886217  u^{24}
\nonumber\\ &&
+11.0163489628560348344708790596  u^{25} + 14.6706757083827785850262439117  u^{26}
\nonumber\\ &&
+19.5975102694203842148444445452  u^{27} + 26.2537143880730842374878981161  u^{28}
\nonumber\\ &&
+35.2639584799959710344981360061  u^{29} + 47.4834621531241956309082413623  u^{30}
\nonumber\\ &&
+64.0845324452896992786471204045  u^{31} + 86.6760348342256170319013958612  u^{32}
\nonumber\\ &&
+117.468474369604443607935643308  u^{33} + 159.502292173007654388803768318  u^{34}
\nonumber\\ &&
+216.963842161082342229133897163  u^{35} + 295.623066963675442979916734791  u^{36}
\nonumber\\ &&
+403.440206590022104677573187301  u^{37} + 551.407438745662532802587646249  u^{38}
\nonumber\\ &&
+754.717250342131567887848557378  u^{39} + 1034.38549262485015775284310232  u^{40} \Biggr] 
\nonumber\\ &&
\times H[0,u]
\nonumber\\ &&
+ \Bigl[
 6.93147180559945309417232121458e\text{-}1    v      -1.93147180559945309417232121458e\text{-}1 v^2
\nonumber\\ &&
+2.64805138932786427505654547915e\text{-}2    v^3    -9.09445606607418535334798246358e\text{-}4 v^4
\nonumber\\ &&
-2.32690966735029895126514843946e\text{-}4    v^5    +2.04869918599655702150619197958e\text{-}5 v^6
\nonumber\\ &&
+3.27244532693809559445428120669e\text{-}6    v^7    -4.76697494718168113031752060736e\text{-}7 v^8
\nonumber\\ &&
-4.79445670857266595292321218748e\text{-}8    v^9    +1.09995268113063922275364870359e\text{-}8 v^{10}
\nonumber\\ &&
+6.30079730899918561829533455058e\text{-}10 v^{11}   -2.4918346059372030398198902277e\text{-}10 v^{12}
\nonumber\\ &&
-5.34007883918865414818687371719e\text{-}12 v^{13}   +5.5156739926803903227757686507e\text{-}12 v^{14}
\nonumber\\ &&
-6.28857786197662796320841934221e\text{-}14 v^{15}   -1.18725105791642065839718279224e\text{-}13 v^{16}
\nonumber\\ &&
+5.26578132632148224808742741623e\text{-}15 v^{17}
+2.46845929257087823714218639074e\text{-}15 v^{18}
\nonumber\\ &&
-1.99041996903111499913398470405e\text{-}16 v^{19}
-4.90370649831330005463993519578e\text{-}17 v^{20}
\nonumber\\ &&
+6.12701864575634669387293564239e\text{-}18 v^{21}
+9.12721047200658523982514560198e\text{-}19 v^{22}
\nonumber\\ &&
-1.70288180301515490338143682483e\text{-}19 v^{23}
-1.5276386825065569828307657806e\text{-}20  v^{24}
\nonumber\\ &&
+4.42944449090124929492159879193e\text{-}21 v^{25}
+2.05134350399689853005153453942e\text{-}22 v^{26}
\nonumber\\ &&
-1.09492124504286623636232861158e\text{-}22 v^{27}
-1.10399957724647642108199227876e\text{-}24 v^{28}
\nonumber\\ &&
+2.58812712884495900131556798953e\text{-}24 v^{29}
-6.10104748471770267083048214269e\text{-}26 v^{30}
\nonumber\\ &&
-5.85427217667691779189384255622e\text{-}26 v^{31}
+3.39895981945488747450890358933e\text{-}27 v^{32}
\nonumber\\ &&
+1.26193521067501965399016009955e\text{-}27 v^{33}
-1.22550544038337779976290204537e\text{-}28 v^{34}
\nonumber\\ && 
-2.56460584430131563394858221159e\text{-}29 v^{35}
+3.76155735706687563080727351239e\text{-}30 v^{36}
\nonumber\\ &&
+4.80146080059348793629186205266e\text{-}31 v^{37}
-1.05400577554603083138457524555e\text{-}31 v^{38}
\nonumber\\ &&
-7.83689686147728007727555462251e\text{-}33 v^{39}
+2.77092887621278377430721664171e\text{-}33 v^{40} \Bigr] 
\nonumber\\ && \times H[0,v]
\end{eqnarray}
for $x \in [0, \sqrt{2} - 1]$.
\begin{table}[H]\centering
\renewcommand{\arraystretch}{1.5}
\begin{tabular}{|r|r|r|r|}
\hline \hline
\multicolumn{1}{|c|}{$x$} 
& \multicolumn{1}{c|}{{\tt HPL}}
& \multicolumn{1}{c|}{run time {\tt Fortran}}
& \multicolumn{1}{c|}{run time {\tt Ginac}}\\
\multicolumn{1}{|c|}{ } 
& \multicolumn{1}{c|}{ }
& \multicolumn{1}{c|}{[sec]}
& \multicolumn{1}{c|}{[sec]}\\
\hline
   $-$20.0 &     5.369919763979762~~   & 2.5e-5  & 0.02923    \\    
           &  $-$18.46370249603318 i   &   &            \\
   $-$0.9  &  $-$1.652038279906588~~   & 2.0e-5  & 0.01917    \\  
           &    +3.344002738868969 i   &   &            \\
   $-$0.2  &  $-$0.067890106575246~~   & 8.6e-5  & 0.00133    \\
           &    +0.068215824899983 i   &      &            \\ 
      0.2  &  $-$0.058464914759637~~   & 5.4e-5  & 0.00189    \\
      0.9  &  $-$0.550223509450311~~   & 8.2e-5  & 0.01310    \\
     50.0  &  $-$18.95831087429180~~   & 3.5e-5  & 0.01260    \\ 
\hline \hline
\end{tabular}
\renewcommand{\arraystretch}{1.0}
\caption[]{\sf Some numerical values for {\tt H[-1,1,0,x]}
and the runtimes in the {\tt FORTRAN} implementation and by {\tt Ginac}.}
\label{TAB2}
\end{table}
\noindent
The following mappings hold~:
\begin{eqnarray}
H[-1,1,0,z] &=& \frac{1}{2} \ln(2) \zeta_2 - \zeta_3 + \zeta_2 H[-1, x] + \frac{1}{6} H[-1, x]^3 - 
 H[-1, x] H[0, -1, x] 
\nonumber\\ &&
- H[-1, x] H[0, 1, x] + H[-1, -1, 1, x] + 
 2 H[0, -1, -1, x] + H[0, -1, 1, x] 
\nonumber\\ &&
+ H[0, 1, -1, x]~~
                                 \text{for}~~z \in [\sqrt{2}-1,1],~~z = \frac{1-x}{1+x}.
\\
H[-1,1,0,-x] &=&
H[-1, x] H[0, -x] H[1, x] - H[0, -x] H[-1, 1, x] - 
 H[1, x] H[0, -1, x] 
\nonumber\\ &&
+ H[0, -1, 1, x]
\\
H\left[-1,1,0,\frac{1}{x}\right] &=&
3 \ln(2) \zeta_2 + \frac{3}{4} \zeta_3 - 2 \zeta_2 H[-1, x] + 2 \zeta_2 H[0, x] - 
\frac{1}{2} H[-1, x] H[0, x]^2 + \frac{1}{6} H[0, x]^3 
\nonumber\\ &&
- H[0, x] H[-1, 1, x] + 
 H[0, x] H[0, -1, x] + H[-1, x] H[0, 1, x] 
\nonumber\\ &&
+ H[0, x] H[0, 1, x] 
- H[0, 0, -1, x] - 2 H[0, 0, 1, x] - H[0, 1, -1, x].
\end{eqnarray}
Here $H[0,-x] = H[0,x] + i \pi$ and one can see that the function is real for $x \geq 0$ and complex
for $x \leq 0$.

\begin{figure}[H] 
\centering
\includegraphics[width=0.48\textwidth]{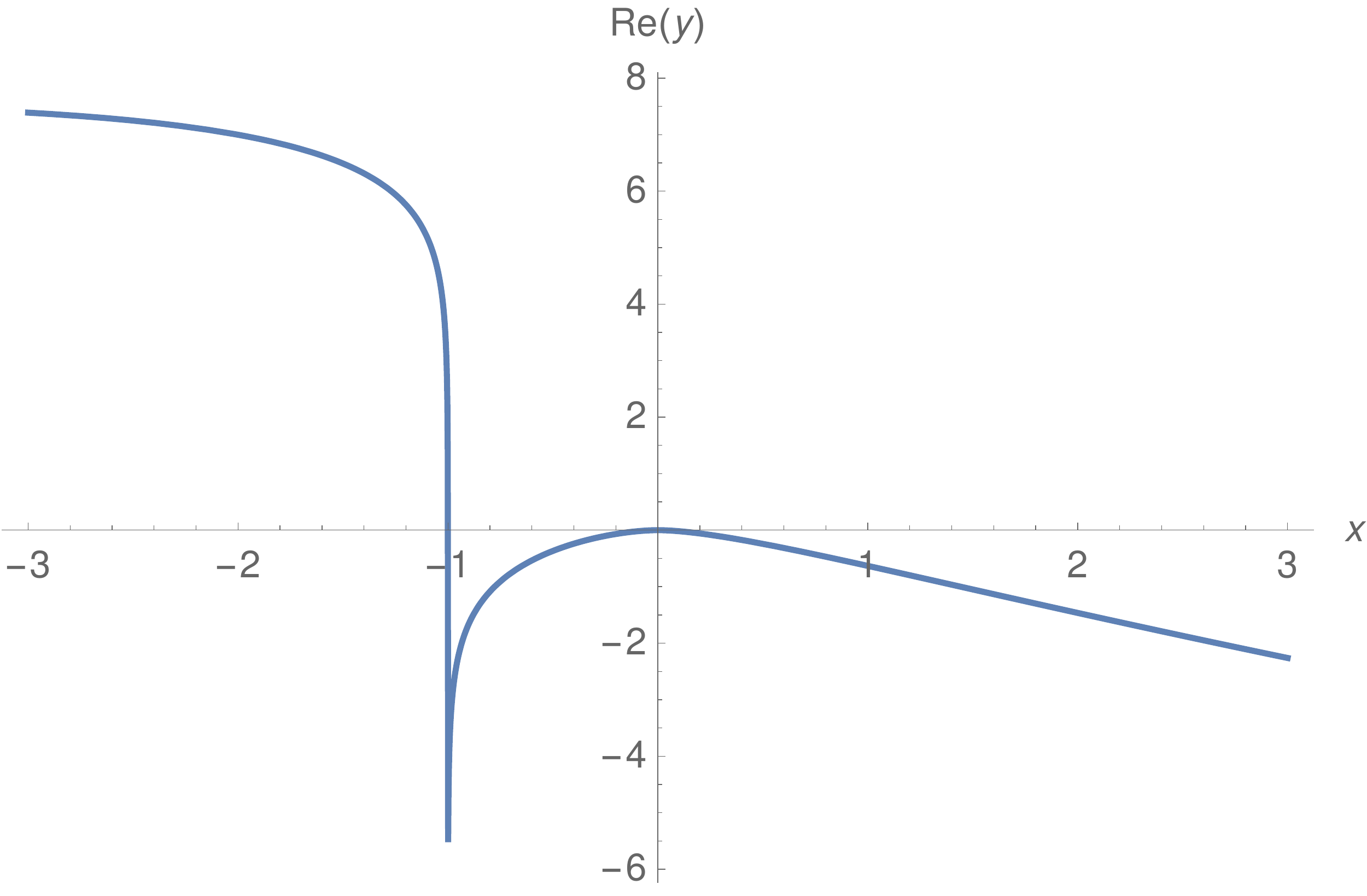}

\vspace*{3mm}
\includegraphics[width=0.54\textwidth]{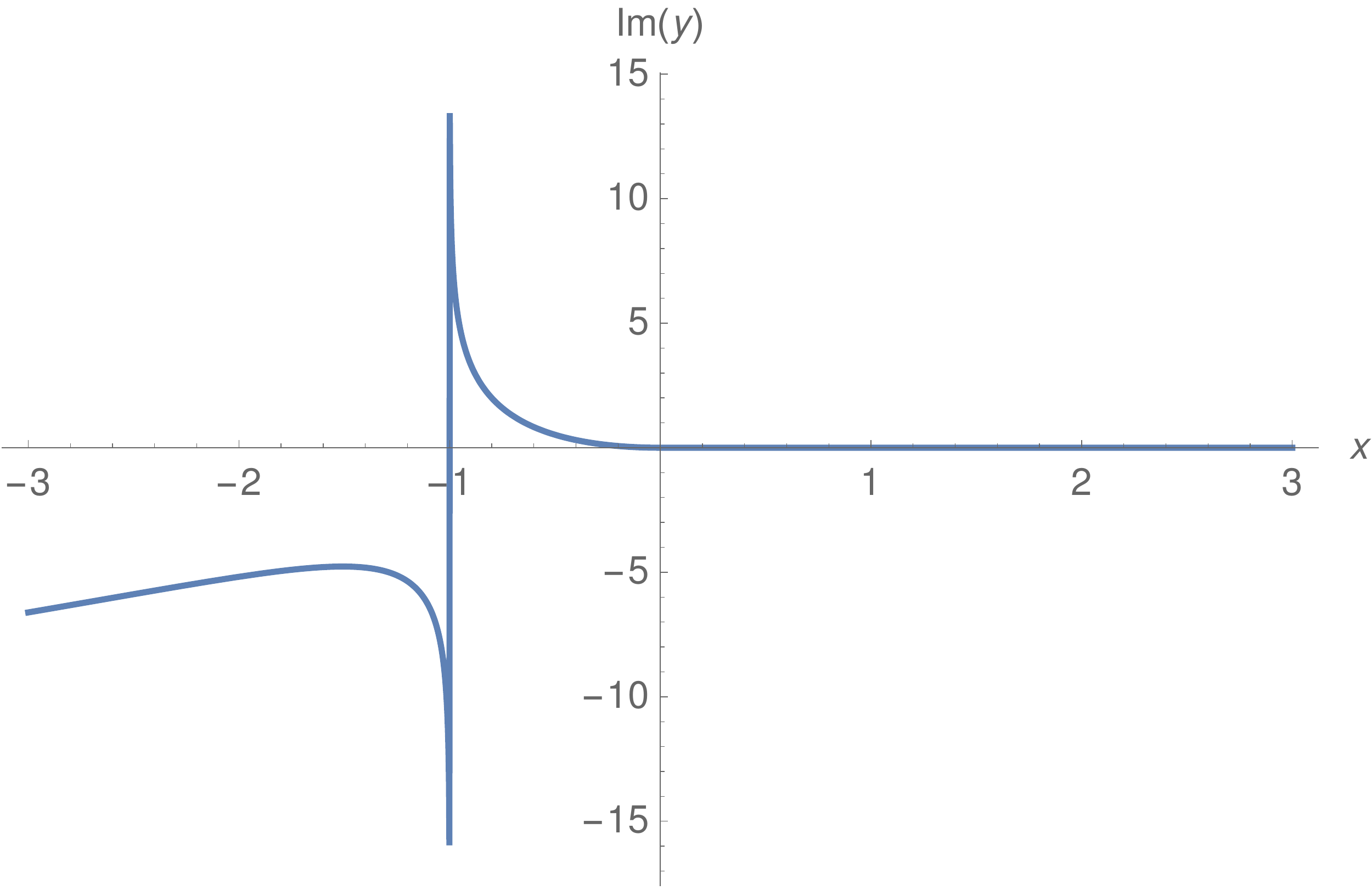}
\caption{\sf \small Real and imaginary part of the function $H[-1,1,0,x]$.}
\label{fig1}
\end{figure}

In Table~\ref{TAB2} we present some numerical results and compare the runtimes of {\tt HPOLY} and {\tt Ginac} 
\cite{Vollinga:2004sn}. In Figure~\ref{fig1} the behaviour of 
$H[-1,1,0,x]$ in the range $x \in [-3,3]$ is illustrated for the real and imaginary part of this function.

We have also measured the run times for all HPLs in {\tt HPOLY} and {\tt Ginac} at $x=0.3$, yielding $9 \cdot 
10^{-3}$ sec and 2.7 sec, respectively.
\subsection{Cyclotomic Harmonic Polylogarithms \label{APPa2}}

\vspace*{1mm}
\noindent
We discuss the representation of the cyclotomic harmonic polylogarithm 
\begin{eqnarray}
{\tt H[\{6,1\},0,-1,x]} = \int_0^x dy_1 \frac{y_1}{1-y_1+y_1^2} \int_0^{y_1} \frac{dy_2}{y_2} 
\int_0^{y_2}
\frac{dy_2}{1+y_2}, 
\end{eqnarray}
i.e. a real 
representation w.r.t. the indices of this iterated integral, and give
some numerical illustrations. The treatment is very similar to the case of the usual harmonic polylogarithms,
since only more letters (regular on $x \in [-1,1]$) are added to the alphabet. In the following 
we display the first 40 expansion coefficients, which show that the series converges only slowly, requiring 
the expansion coefficients $|\xi| \leq 1/2$ to reach double precision. 
One obtains
\begin{eqnarray}
\lefteqn{{\tt H[\{6,1\},0,-1,x]}}   \nonumber\\ &&
0.333333333333333333333333333333 x^3+0.187500000000000000000000000000 x^4
\nonumber\\ &&
-0.0277777777777777777777777777778 x^5-0.158564814814814814814814814815 x^6
\nonumber\\ &&
-0.110357142857142857142857142857 x^7+0.0188888888888888888888888888889 x^8
\nonumber\\ &&
+0.104891030486268581506676744772 x^9+0.0777283163265306122448979591837 x^{10}
\nonumber\\ &&
-0.0140354938271604938271604938272 x^{11}-0.0784727996136726295456454186613 x^{12}
\nonumber\\ &&
-0.0599245154196452897751599050300 x^{13}+0.0111221750331598816447301295786 x^{14}
\nonumber\\ &&
+0.0627097540455293813047170800529 x^{15}+0.0487396137126484903707680930458 x^{16}
\nonumber\\ &&
-0.00919812040212473112906013338914 x^{17}-0.0522282286799164597292395420194 x^{18}
\nonumber\\ &&
-0.0410673611468378472909181683198 x^{19}+0.00783709173477453784228900692598 x^{20}
\nonumber\\ &&
+0.0447519895041858538174032906623 x^{21}+0.0354795429496550987872585893358 x^{22}
\nonumber\\ &&
-0.00682496786953676661431421089295 x^{23}-0.0391495967330932143454954592724 x^{24}
\nonumber\\ &&
-0.0312290280570659818475651680142 x^{25}+0.00604340419447986488429425811813 x^{26}
\nonumber\\ &&
+0.0347946003882639271548437004232 x^{27}+0.0278873718387793265650809033939 x^{28}
\nonumber\\ &&
-0.00542193299602834607361795351938 x^{29}-0.0313119326191574937536749057194 x^{30}
\nonumber\\ &&
-0.0251913826144454848304917188798 x^{31}+0.00491606270049386474230914922487 x^{32}
\nonumber\\ &&
+0.0284632560663618316507637169020 x^{33}+0.0229704965080628462565091141932 x^{34}
\nonumber\\ &&
-0.00449635129343592282670836238196 x^{35}-0.0260898396790974516252031566278 x^{36}
\nonumber\\ &&
-0.0211093407201794247209341019341 x^{37}+0.00414255307533152470774948058168 x^{38}
\nonumber\\ &&
+0.0240818739410040077679013044162 x^{39}+0.0195270886515233773838902417075 x^{40}
\nonumber\\
\end{eqnarray}
for $|x| \leq 1/2$,  
\begin{eqnarray}
\lefteqn{{\tt H[\{6,1\},0,-1,x]} =}   \nonumber\\ &&
 0.340390207765402505097420897861 - 0.822467033424113218236207583323 x_- 
\nonumber\\ &&
+0.346573590279972654708616060729 x_-^2  + 0.306346874568028624314941214684 x_-^3 
\nonumber\\ &&
+0.0380088949293707529895132089211 x_-^4 - 0.156620128473319660687641363698 x_-^5 
\nonumber\\ &&
-0.156597541296457541255631321152 x_-^6 - 0.0225554460774175569791333271786 x_-^7 
\nonumber\\ &&
+0.0976519307918251785767956228722 x_-^8 + 0.104325369067180060242342866096 x_-^9 
\nonumber\\ &&
+0.0157646326300601883740158725548 x_-^{10} - 0.0710280097948583028355514206622 x_-^{11}
\nonumber\\ &&
-0.0782470790060635646220138334389 x_-^{12} - 0.0121277817768820409689235610102 x_-^{13}
\nonumber\\ &&
+0.0558072840567786665216825950946 x_-^{14} + 0.0625974915019450287029214667407 x_-^{15} 
\nonumber\\ &&
+0.00985375412765262691406233891207 x_-^{16} - 0.0459589676261932861599464151009 x_-^{17} 
\nonumber\\ &&
-0.0521645876903227753783336731631 x_-^{18} - 0.00829790297360334455491929168585 x_-^{19} 
\nonumber\\ &&
+0.0390651204796707863269908727563 x_-^{20} + 0.0447125028842087830347950513173 x_-^{21} 
\nonumber\\ &&
+0.00716637038578137582472531686648 x_-^{22} - 0.0339696701394558426216208414143 x_-^{23}
\nonumber\\ &&
-0.0391234400919570434982680106204 x_-^{24} - 0.00630640596933985512905044780475 x_-^{25} 
\nonumber\\ &&
+0.0300500928025446660489690952202 x_-^{26} + 0.0347763911870470964246495341404 x_-^{27} 
\nonumber\\ &&
+0.00563071961290383166332573519540 x_-^{28} - 0.0269414625137767299362368745328 x_-^{29} 
\nonumber\\ &&
-0.0312987520688579112914323618433 x_-^{30} - 0.00508581126349987627019732720481 x_-^{31} 
\nonumber\\ &&
+0.0244157004030054693980770411665 x_-^{32} + 0.0284534109716415871863511989614 x_-^{33} 
\nonumber\\ &&
+0.00463706321081657316300068964177 x_-^{34} - 0.0223229260827576788486381082623 x_-^{35} 
\nonumber\\ &&
-0.0260822933906759433489536861150 x_-^{36} - 0.00426108511264431308690488692722 x_-^{37} 
\nonumber\\ &&
+0.0205605898130653384249909302377 x_-^{38} + 0.0240759631298542819670313546736 x_-^{39} 
\nonumber\\ &&
+ 0.00394150372919578889368783808396 x_-^{40}
\end{eqnarray}
for $x \in [1/2,1]$ and
\begin{eqnarray}
\lefteqn{{\tt H[\{6,1\},0,-1,x]} =}   \nonumber\\ &&
-0.194568246286013745578625203082
\nonumber\\ &&
+\Bigl(-0.548311355616075478824138388882     x_+ 
+0.250000000000000000000000000000     x_+^2
\nonumber\\ &&
+0.107219780253638016165645006172     x_+^3
+0.271740944494877713834930138883e\text{-}1  x_+^4
\nonumber\\ &&
-0.257505086150775649670496049413e\text{-}2  x_+^5
-0.943456337336485738803035905363e\text{-}2  x_+^6
\nonumber\\ &&
-0.816905052458110731788689690942e\text{-}2  x_+^7
-0.520795277987738305198622658940e\text{-}2  x_+^8
\nonumber\\ &&
-0.277060313984829741036690222526e\text{-}2  x_+^9
-0.126851938027123900831848220794e\text{-}2  x_+^{10}
\nonumber\\ &&
-0.503667726779323063780400849847e\text{-}3  x_+^{11}
-0.180235498282985631923185530178e\text{-}3  x_+^{12}
\nonumber\\ &&
-0.733668756985816187679438098227e\text{-}4  x_+^{13}
-0.517342767257388203755181868192e\text{-}4  x_+^{14}
\nonumber\\ &&
-0.529768547494945595363266938968e\text{-}4  x_+^{15}
-0.541523823048175872246118351091e\text{-}4  x_+^{16}
\nonumber\\ &&
-0.504956623487350579432257136818e\text{-}4  x_+^{17}
-0.435057422798886284192642625177e\text{-}4  x_+^{18}
\nonumber\\ &&
-0.355976912750893316702486185276e\text{-}4  x_+^{19}
-0.283747696506692566232934066468e\text{-}4  x_+^{20}
\nonumber\\ &&
-0.224862637086975580231554304003e\text{-}4  x_+^{21}
-0.179654244130490719256568063724e\text{-}4  x_+^{22}
\nonumber\\ &&
-0.145743279416543605609014732902e\text{-}4  x_+^{23}
-0.120215246792584256329867181528e\text{-}4  x_+^{24}
\nonumber\\ &&
-0.100603516738056680791340289733e\text{-}5  x_+^{25}
-0.851341006823932654661521938303e\text{-}5  x_+^{26}
\nonumber\\ &&
-0.726349827322799151408472485058e\text{-}5  x_+^{27}
-0.623549500679759632941394636300e\text{-}5  x_+^{28}
\nonumber\\ &&
-0.538028643232232014650606829135e\text{-}5  x_+^{29}
-0.466381278035087853538774282359e\text{-}5  x_+^{30}
\nonumber\\ &&
-0.406070457715978200292192742285e\text{-}5  x_+^{31}
-0.355097711186231539907222170265e\text{-}5  x_+^{32}
\nonumber\\ &&
-0.311840966987030702179107122442e\text{-}5  x_+^{33}
-0.274971225781815571600317219459e\text{-}5  x_+^{34}
\nonumber\\ &&
-0.243400658296993141391406093405e\text{-}5  x_+^{35}
-0.216242026558948610498083647252e\text{-}5  x_+^{36}
\nonumber\\ &&
-0.192773190450817788358838757683e\text{-}5  x_+^{37}
-0.172405883754813902219634478337e\text{-}5  x_+^{38}
\nonumber\\ &&
-0.154659185525022548769109014074e\text{-}5  x_+^{39}
-0.139137829546522993754067930066e\text{-}5  x_+^{40}
\nonumber\\ &&
+\Bigl[
-0.166666666666666666666666666667 x_+^2
-0.555555555555555555555555555556e\text{-}1 x_+^3
\nonumber\\ &&
+0.166666666666666666666666666667e\text{-}1 x_+^5
+0.166666666666666666666666666667e\text{-}1 x_+^6
\nonumber\\ &&
+0.119047619047619047619047619048e\text{-}1 x_+^7
+0.724206349206349206349206349206e\text{-}2 x_+^8
\nonumber\\ &&
+0.401234567901234567901234567901e\text{-}2 x_+^9
+0.214285714285714285714285714286e\text{-}2 x_+^{10}
\nonumber\\ &&
+0.119047619047619047619047619048e\text{-}2 x_+^{11}
+0.748556998556998556998556998557e\text{-}3 x_+^{12}
\nonumber\\ &&
+0.549450549450549450549450549451e\text{-}3 x_+^{13}
+0.448955806098663241520384377527e\text{-}3 x_+^{14}
\nonumber\\ &&
+0.382395382395382395382395382395e\text{-}3 x_+^{15}
+0.326756576756576756576756576757e\text{-}3 x_+^{16}
\nonumber\\ &&
+0.276765717942188530423824541472e\text{-}3 x_+^{17}
+0.232655869910771871556185281675e\text{-}3 x_+^{18}
\nonumber\\ &&
+0.195199537304800462695199537305e\text{-}3 x_+^{19}
+0.164375742936114453142316919407e\text{-}3 x_+^{20}
\nonumber\\ &&
+0.139449719281652054761298458777e\text{-}3 x_+^{21}
+0.119375300560214717265097225694e\text{-}3 x_+^{22}
\nonumber\\ &&
+0.131134477762556911836761076230e\text{-}3 x_+^{23}
+0.897896648502382786943293067940e\text{-}4 x_+^{24}
\nonumber\\ &&
+0.787312103101576785787312103102e\text{-}4 x_+^{25}
+0.694430190195227419411231291907e\text{-}4 x_+^{26}
\nonumber\\ &&
+0.615646514102514860305899426862e\text{-}4 x_+^{27}
+0.548299963598119473742033978943e\text{-}4 x_+^{28}
\nonumber\\ &&
+0.490370738953877725049414746362e\text{-}4 x_+^{29}
+0.440279180491117866083633973683e\text{-}4 x_+^{30}
\nonumber\\ &&
+0.396759437506466368945146534281e\text{-}4 x_+^{31}
+0.358780629639696323771643547090e\text{-}4 x_+^{32}
\nonumber\\ &&
+0.325495384282835477024478780285e\text{-}4 x_+^{33}
+0.296203528090464701585299046187e\text{-}4 x_+^{34}
\nonumber\\ &&
+0.270324458622603140106269524853e\text{-}4 x_+^{35}
+0.247375088378105046141745858401e\text{-}4 x_+^{36}
\nonumber\\ &&
+0.226951904770740284572302948414e\text{-}4 x_+^{37}
+0.208716366275858749420441086658e\text{-}4 x_+^{38}
\nonumber\\ &&
+0.192383083866313435144166242061e\text{-}4 x_+^{39}
+0.177710310170514453671986061289e\text{-}4 x_+^{40}
\Bigr] 
\nonumber\\ && 
\times H[-1,x]\Bigr)
\end{eqnarray}
for $x \in [-1,-1/2]$. Here we used the notation $x_{\pm} = (1 \pm x)$.

\section{List of the attached files \label{APPb}}
The following files are attached to this paper. The analytic files are given in {\tt Mathematica} 
format and can be easily converted into {\tt Maple} or {\tt FORM} \cite{Vermaseren:2000nd} format.
\begin{table}[H]\centering
\renewcommand{\arraystretch}{1.5}
\begin{tabular}{|r|l|}
\hline \hline
\multicolumn{1}{|c|}{File name} 
& \multicolumn{1}{c|}{contents}
\\
\hline
   {\tt LISTn.m}  &  Bases per weight {\tt n}, {\tt n = 1...8}\\
   {\tt LIST.m}   &  Total basis \\
   {\tt hrel.m}   &  Replacement rules to basis for all {\tt HPLs} up to weight {\sf w=8}.    \\
   {\tt LISTOMX.m}  &  Replacement rules: $x \mapsto -x$ for the basis elements \\ 
   {\tt LISTOOX.m}  &  Replacement rules: $x \mapsto 1/x$ for the basis elements \\
   {\tt LISTMTX.m}  &  Replacement rules: $x \mapsto (1-x)/(1+x)$ for the basis elements \\
\hline 
   {\tt HPOLY.f}  &  {\tt FORTRAN} code~~(1298 lines)\\
   {\tt TTTn.m}     &  Data files containing the numerical constants needed at the different \\
                  &  weights,~ {\tt n = 2...8}. 
the different weights \\
   {\tt Bkl.m}    &  Data files to identify the basis elements: {\tt k=1 .. k=l, k = 5,6,7,8} 
\\
\hline \hline
\end{tabular}
\renewcommand{\arraystretch}{1.0}
\label{TAB4}
\caption[]{\sf The list of attached files 
}
\end{table}

\noindent
The data files {\tt TTTn.m} contain 360, 1600, 5040, 17280, 51040, 162240, 486000 constants
for {\sf w = 2...8}. Initially the external files for reading or writing are all set to {\tt STATUS="NEW"}, as e.g.

\begin{verbatim}
      OPEN(UNIT=14,FILE="TTT2.m",FORM="FORMATTED",STATUS="NEW",
     &     ACTION="READ") 
\end{verbatim}

\noindent
In later use the standard input files have to be set to {\tt STATUS="OLD"}.
For convenience we provide two versions of {\tt HPOLY.f}. {\tt HPOLYnew.f} contains the file status: {\tt "NEW"} and 
{\tt HPOLY.f} the file status {\tt "OLD"}.

\vspace*{5mm}
\noindent
{\bf Acknowledgment.} We would like to thank R.~Germundson, P.~Marquard, N.~Rana, and K.~Sch\"onwald 
for discussions and N.~Rana for an introduction to and help with using {\tt Ginac}. This work was 
supported in part by the Austrian Science Fund (FWF) grant SFB F50 (F5009-N15), 
by the bilateral project DNTS-Austria 01/3/2017 (WTZ BG03/2017), funded by the Bulgarian National 
Science Fund and OeAD (Austria), by the EU TMR network SAGEX Marie Sk\l{}odowska-Curie grant agreement No. 764850  
and COST action CA16201: Unraveling new physics at the LHC through the precision frontier.  
M.~Round thanks DESY and the KMP Berlin for support. 


\end{document}